# Correction to 'Equations of turbulent motion of an incompressible fluid (A.N. Kolmogorov)' (in English, Translation by D.B. Spalding). *Proc. R. Soc. Lond.* **A434**, 214-216 (1991).

John Z. Shi

State Key Laboratory of Ocean Engineering,
School of Naval Architecture, Ocean and Civil Engineering,
Shanghai Jiao Tong University,
1954 Hua Shan Road, Shanghai 200030, China
zshi@sjtu.edu.cn

Kolmogorov published his famous $b - \omega$ model in 1942 (Kolmogorov 1942) when the German wehrmacht was already deep in Soviet territory, i.e. so called the Kolmogorov's two-equation of turbulence, or the Kolmogorov's $k - \omega$ two-equation turbulence closure model. D.B. Spalding translated it into English (Kolmogorov 1991s)(s for the first letter of Spalding) and commented on it (Spalding 1991). Two important physical quantities are introduced in Kolmogorov (1942, 1991): i) "the rate of dissipation of energy in unit volume and time" ($w$); ii) "some mean 'frequency' ($\omega$)". Kolmogorov (1991, page 214, paragraph 1, lines 11-12) highlighted the quantity i) [i.e., $w$] as 'A fundamental characteristic of the turbulent motion at all scales'. However, there are four typos appeared in the English version (Kolmogorov 1991), among which three typos are related to the same symbol '$w$' being misprinted as the Greek letter '$\omega$'. They are misleading and can cause severe confusion/misinterpretation/misunderstanding.

As the first example, in his own interpretation, Spalding (1991, page 212, paragraph 2) wrote that

> The variables which Kolmogorov chose for the characterization of turbulence were the fluctuation energy ($b$……) and the frequency ($\omega$, in his notation…)…and the second, if multiplied by $b$, is proportional to the energy-dissipation rate (usually given the symbol $\varepsilon$, in contrast to Kolmogorov's use of the same symbol [$\omega$]…..

As the second example, in his famous review paper, Wilcox (1991, page 1, right column, bottom Section/paragraph) wrote that

> A. Kolmogorov's Model
> 
> The very first two-equation turbulence model was formulated by Kolmogrov[5]. He referred to $\omega$ as "the rate of dissipation of energy in unit volume and time." To underscore its physical relation to the "'external scale' of turbulence, $L$" he also called it "some mean 'frequency' determined by $\omega = ck^{\frac{1}{2}}/L$, where is $c$ a constant." [Note that Kolmogorov (1942, 1991) used the letter '$b$' instead of '$k$': $\omega = cb^{\frac{1}{2}}/L$]

Another related misprint is the symbol "$w$" used in Bulicek and Malek (2019, page 105, paragraph 3, line 5).

After carefully checking the original Russian version of Kolmogorov (1942), in the present author's view, both Spalding's and Wilcox's misinterpretations are due to the fact that they did not notice the two different symbols, '$w$' and '$\omega$', which Kolmogorov (1942) actually used.

The present author believes that there are other similar misinterpretations in the literature. To avoid them, the present author feels that the following necessary corrections should be made to Kolmogorov (1991):

Page 214, paragraph 1, line 12, *for* $\omega$ *read* $w$

Page 215, paragraph 1, line 9, *for* $v^{\frac{3}{4}}(\rho\omega)^{\frac{1}{4}}$ *read* $v^{\frac{3}{4}}(\rho/w)^{\frac{1}{4}}$. Please note that two typos: the slash '/' and the symbol '$w$'

Page 215, paragraph 3, line 10 below equation (3), *for* $(\frac{2}{3}\omega/\rho)$ *read* $(\frac{2}{3}w/\rho)$

In a way, these corrections will alert those who can read Russian to the two different symbols, '$w$' and '$\omega$', which Kolmogorov (1942) used in the first place.

In addition, two more typos should be corrected to Kolmogorov (1991):

Page 215, paragraph 1, line 4, *for* $B_{nn}$ and $B_{tt}$ *read* $B_{dd}$ and $B_{nn}$
Page 215, paragraph 1, line 7, *for* $B_{nn}$ and $B_{tt}$ *read* $B_{dd}$ and $B_{nn}$

Unlike those typos found earlier, these two more typos do not significantly affect the understanding of Kolmogorov (1991s).

One may argue that it could be pointed out that in the literature these days it is relatively common to see turbulent energy dissipation rate per unit mass (often denoted "$\epsilon$") rather than Kolmogorov's "$w$" which was turbulent energy dissipation rate per unit volume. The quantity $\rho/w$ appearing at the end of the manuscript is actually $1/\epsilon$. If researchers are using "$\epsilon$" instead of "$w$" the risk of confusing it with "$\omega$" may be less. However, this author believes that Kolmogorov used the symbol $w$ because of his own solid reason/argument. As Kolmogorov (1991s, page 214, paragraph 1, lines 11-13) wrote

> A fundamental characteristic of the turbulent motion at all scales is the quantity $\omega$ [$w$] which stands for the rate of dissipation of energy in unit volume and time.

Nevertheless, Kolmogorov actually used the ratio of $w$ to the density ($\rho$) throughout. We can believe that such a ratio seems to him more justified from a physical point of view. Indeed, the value of the average energy dissipation rate per unit volume is obtained as a result of dividing the integral energy dissipation rate in a certain volume by the value of this volume. This results in a value ($w$) that, as Kolmogorov highlighted, is the same for all length scales. If the averaging is made not by the volume, but, instead, by the mass, then it is no longer possible to talk about the invariance of the corresponding averaged value for all length scales. Only in the case when the density ($\rho$) is constant, the result must be the same. Of course, Kolmogorov did consider the case of incompressible medium.

The dimensions of $w$ are {[Mass]·[Length]$^2$·[Time]$^{-2}$ [Length]$^{-3}$·[Time]$^{-1}$} or {[Mass]·[Length]$^{-1}$·[Time]$^{-3}$} whereas the dimensions of $\omega$ are [Time]$^{-1}$. Hopefully, a reader will be able to negotiate their way through Kolmogorov's work merely track of the dimensions of quantities.

## Acknowledgements

Xian-Liang Gong first drew the present author's attention to 'the rate of dissipation of energy in unit volume and time' in Kolmogorov (1991). Sergey Chefranov is thanked for finding the author the original Russian Kolmogorov (1942) and having drawn my attention to the misprint of $w$.